\documentclass[5p,times,twocolumn]{elsarticle}

\usepackage[english]{babel}
\usepackage{hyperref}
\usepackage{amsfonts} 
\usepackage{amsmath}
\usepackage{amssymb}
\usepackage{graphicx}
\usepackage{color}
\usepackage{natbib}  
\usepackage{bm} 
\usepackage{isomath} 
\usepackage{braket}
\usepackage{fleqn}
\usepackage{txfonts}

\newcommand{\R}{\mathbb{R}}

\begin{document}

\begin{frontmatter}
\title{Wigner instability analysis of the damped Hirota equation} 

\author[1]{Al--Tarazi Assaubay}

\author[1]{Alejandro J. Castro}
\address[1]{Department of  Mathematics, Nazarbayev University,
		010000 Nur-Sultan, Kazakhstan}

\author[2]{Antonio A. Valido}
\ead{a.valido@iff.csic.es}
\address[2]{Instituto de F\'isica Fundamental IFF-CSIC, Calle Serrano 113b, 28006 Madrid, Spain}

\date{\today}

\begin{abstract}

We address the modulation instability of the Hirota equation in the presence of stochastic spatial incoherence and linear time-dependent amplification/attenuation processes via the Wigner function approach. We show that the modulation instability remains baseband type, though the damping mechanisms substantially reduce the unstable spectrum independent of the higher-order contributions (e.g.  the higher-order nonlinear interaction and the third-order dispersion). Additionally, we find out that the unstable structure due to the Kerr interaction exhibits a significant resilience to the third-order-dispersion stabilizing effects in comparison with the higher-order nonlinearity, as well as a moderate Lorentzian spectrum damping may assist the rising of instability.  Finally, we also discuss the relevance of our results in the context of current experiments exploring extreme wave events driven by the modulation instability (e.g. the generation of the so-called rogue waves).  
\end{abstract}


\end{frontmatter}

\section{Introduction}

During the last decade there has been an important theoretical and experimental progress in the generation as well as manipulation of extreme wave events which are governed by the nonlinear Schr\"odinger equation (NLSE). For instance, Refs. \cite{chabchoub20111,peric20151,kibler20101,bailung20111} recently reported the experimental observation of the Peregrine soliton in surface gravity wave, ultrashort light pulse, and multicomponent plasma scenarios. This is the simplest solution of the NLSE which is localized in both space and time, and it is regarded as a prototype of the so-called rogue waves \cite{chen20171,onorato20131,dudley20141}. Another genuinely nonlinear feature observed in these experimental platforms was the Fermi-Pasta-Ulam-Tsingou (FPUT) recurrence \cite{kimmoun20161,simaeys20011}, which corresponds to the breaking of a continuous wave into a periodic train of spatially localized wave packets \cite{dudley20091}. Although there is no experimental evidence yet (to the best of our knowledge), these "freak" wave phenomena has been theoretically predicted to occur in higher-order extensions of the NLSE as well \cite{ankiewicz20101,ankiewicz20171,tao20121,li20131,ankiewicz20141}, such as the Hirota equation \cite{hirota19731}. The latter equation is indispensable for an accurate description of the nonlinear wave dynamics when the third-order dispersion or delayed nonlinear response are significant, for instance in the propagation of femtosecond pulses in highly dispersive optical fibers \cite{hao20041,mahalingam20011}. 

Today it is broadly accepted that the (coherent) modulation instability (MI) is one of the principal physical mechanisms explaining the rise of amplitude-growing periodic perturbations from unstable quasi-continuous pulses \cite{kibler20151,picozzi20141,zakharov20091} (see also \cite{sotocrespo20161}). For instance, the baseband-type MI  has been shown to be fundamental for the formation of rogue waves in the NLSE 
\cite{dudley20091,
kibler20151,
akhmediev19871,
akhmediev20091,
onorato20091,
guo20161,
chabchoub20151}, 
the Hirota equation \cite{ankiewicz20101,xu20031}, and others higher-order generalized equations \cite{baronio20151,baronio20141,chen20151,achilleos20151,yan20131,wang20161,zhang20171,zhang20191,shukla20051,marklund20061}. In this line, in \cite{chowdury20181} the authors have recently shown that the higher-order non-linear effects substantially influence the formation of the so-called Akhmediev breathers (AB), which may eventually lead to extreme wave events \cite{akhmediev20091}.

Unfortunately, the MI mechanism is fragile to the unavoidable dissipation and noise (which commonly render stabilizing effects) present in most interesting situations, so a complete understanding of the experimental observation of extreme wave events driven by the MI starts by addressing the influence of realistic damping effects. In this line, it is only recently that the (coherent) MI dynamics has been extensively analysed for the case of the NLSE subjected to (linear) gain/loss effects in drifted/dissipative surface gravity waves \cite{
kimmoun20161,
segur20051,
kimmoun20171,
athanassoulis20171,
onorato20121} 
and ultrashort pulses propagating in amplified/attenuated optical fibers 
\cite{
hao20041,
yan20131,
kraych20191,
sagnes20101}. 
Another leading-order effect coming from experimental imperfections (e.g. incoherent spatial/temporal light sources \cite{sun20121}) is the existence of phase correlations due to partially incoherent pulses 
\cite{
picozzi20141,
kibler20121}. 
This leads to the so-called incoherent MI dynamics \cite{soljanic20001}, and importantly, it closely resemblances the well-known (energy-conserving) Landau damping from electron plasma waves \cite{fedele20001,hall20021,hall20041,helczynski20031}. For instance, the incoherent effects are behind the formation of incoherent solitons in nonlinear optics \cite{kibler20121,kip20001}. Although this damping mechanism has found to have an equivalent interpretation to the famous Landau damping in the context of surface gravity waves in deep waters \cite{onorato20031,athanassoulis20201}, this is not the case in the more general scenario treated here, so it shall be refereed to as Lorentzian spectrum damping to distinguish from this. Interestingly, the Lorentzian spectrum damping was originally found out via the application of the celebrated Wigner-function framework in the MI analysis of the NLSE 
\cite{
fedele20001,
hall20021,
onorato20031,
fedele20002}. 
These works further motivated the use of the Wigner-function approach to address the MI in interacting or higher-order nonlinear equations subjected to incoherent sources 
\cite{
shukla20051,
marklund20061,
hall20041,
helczynski20031,
anderson20041}, as well as the formation and dynamics of incoherent wave solutions \cite{picozzi20141}. Nonetheless, the Wigner function was previously used to study the instability of the NLSE in presence of a random background \cite{albert19781} or four-wave interactions in the realm of surface gravity waves \cite{janssen20031}, and most recently, it has been employed as a route to apply the Penrose instability analysis to the study of rogue waves \cite{athanassoulis20171,athanassoulis20201}. Despite the preceding efforts to understand the impact of these damping effects, the competition between them and the higher-order nonlinear features from the Hirota equation is still largely unexplored.

In the present work we investigate the action of both the Lorentzian spectrum damping and the linear gain/loss contribution upon the MI characteristic of the one-dimensional Hirota equation 
(note the subscript $t$ and $x$ denote the partial derivatives in the propagating and transverse variables, respectively) \cite{ankiewicz20101,tao20121},
\begin{equation}\label{eq:HE}
i  \psi_t + i \alpha |\psi|^2  \psi_x + \beta  \psi_{xx} + i \gamma \psi_{xxx} + \theta |\psi|^2 \psi=i\eta(t)\psi,
\end{equation}
where $\psi$ stands for the usual wave function envelope (e.g. modulating either the electric field in an optical fiber \cite{ankiewicz20101} or the surface elevation in deep water \cite{tao20121}), $\beta$ is the group velocity dispersion (GVD), $\theta$ determines the Kerr nonlinearity or self-phase modulation, and $\eta(t)$ characterizes a time-dependent amplification/attenuation process. Adopting the denomination suggested in 
\cite{
mahalingam20011,
chen20131}, 
the terms $\gamma$ and $\alpha$ shall be referred to as the third-order dispersion (TOD) and self-steepening (SS), respectively. By following the incoherent MI analysis in the celebrated Wigner-function framework  
\cite{
athanassoulis20171,
hall20041,
onorato20031}, we show that the Kerr interaction is more resilient to the third-order-dispersion stabilizing effects than the higher-order nonlinear interaction. We also see the SS contribution in conjunction with the Lorentzian spectrum damping may augment the unstable spectrum
in agreement with previous works dealing with generalized NLSEs \cite{shukla20051,marklund20061}. Nonetheless any (first-order) instability signature eventually disappears by the action of strong damping strengths. Overall, our investigation sheds new light about the interplay between higher-order contributions to the nonlinear dynamics and both damping mechanism for a broad set of experimental parameters \cite{fedele20002}, and it suggests that the aforementioned wave events in the context of the Hirota equation are experimentally amenable with the current technology in either optical fiber or water tank experiments under feasible damping conditions. Let us emphasize that our work substantially differs from the vast majority of previous treatments using the Wigner-function framework because they restrict themselves to study separately the formation of MI in the NLSE under the influence of either the Lorentzian spectrum damping \cite{onorato20131,fedele20001,fedele20002,janssen20031} or dissipative effects 
\cite{
shukla20051,
marklund20061,
athanassoulis20171,
coppini20191}. In contrast, we provide a unified framework that ultimately returns known previous results as particular instances (e.g., we recover the instability growth rate from the NLSE \cite{soljanic20001}), as well as we deepen in the interplay between non-linearity effects and damping mechanisms. Furthermore, our motivation is different from previous works, this stems in the observation that higher-order non-linear effects may be non-trivially perturbed by the damping processes.

The present paper is organized as follows. In Sec.\ref{WIA} we start illustrating the standard procedure to analyze the modulation instability of the Hirota equation in the framework of the Wigner function in absence of amplification/attenuation processes (i.e. $\eta(t)=0$), as well as we introduce the definition of the (first-order) modulation instability and the unstable spectrum. This section also displays the generalized dispersion relation which is the basis of the subsequent section. In Sec.\ref{Sec:BF} we study the well-known Benjamin-Feir instability, whereas both numerical and analytical results from the incoherent MI analysis are presented for a broad class of problem parameters in Sec.\ref{Sec:IMI}. For the seek of clarity, the latter analysis is separately carrying out for vanishing ($\gamma=0$) and non-vanishing ($\gamma\neq 0$) TOD effects in Secs. \ref{Sec:IMI1} and \ref{Sec:IMI2}, respectively. Then, the impact of linear loss/gain contributions (i.e. $\eta(t)\neq 0$) to the MI is fully addressed in Sec.\ref{Sec:LD}. In Sec.\ref{EOEWE} all these results are further discussed in the context of two feasible experimental platforms: ultrashort pulse in optical fibers and surface gravity waves in water tanks. Finally, we summarize and draw the main conclusions in Sec.\ref{Sec:CR}.
 
\section{Wigner instability analysis without gain/loss effects}\label{WIA}

Before presenting our findings in presence of gain/loss effects, for the seek of clarity it is convenient to briefly refresh the MI analysis in the Wigner framework in the simpler case of disregarding dissipative effects
\cite{
hao20041,
picozzi20141,
fedele20001,
onorato20031,
fedele20002}. Let us thus assume a vanishing gain/loss rate (i.e. $\eta(t)=0$), which shall be treated in full detail in Sec.\ref{Sec:LD}. Instead of carrying out our treatment in terms of the  wave function $\psi$, we switch to the so-called Wigner function, denoted by $W$, via the Wigner-Moyal transform \cite{hall20021,helczynski20031,onorato20031}, which is given by
\begin{equation}
 W(x,k,t) = \int_{\R} e^{- i k y} \, \psi\Big(x+\frac{y}{2},t\Big) \, \psi^{*} \Big(x-\frac{y}{2},t\Big) dy, 
 \label{eq:WignerTransf}
\end{equation}
where $k \in \R$ represents the usual wave number coordinate, and $\psi^{*}$ denotes the complex conjugate. In the Wigner-Moyal description, the Hirota equation (\ref{eq:HE}) governing the wave-function dynamics 
reads

\begin{align}\label{eq:HirotaWigner}
& W_t
 =  \Big(- 2\beta k + 3\gamma k^2 \Big)\, W_x 
- \frac{\gamma}{4}  \, W_{xxx} \nonumber \\
& + \frac{1}{2\pi} \int_{\R}\int_{\R} \Big(i \theta -i \alpha (k-\lambda) \Big)e^{-i \lambda y} 
\Big\{ Z\Big(x+\frac{y}{2},t\Big) - Z\Big(x-\frac{y}{2},t\Big) \Big\}\, dy \nonumber \\
& \qquad \qquad \times W(x,k-\lambda,t) d\lambda \nonumber\\
& -  \frac{\alpha}{4\pi} \int_{\R}\int_{\R}  e^{-i \lambda y} \, 
\Big\{ Z\Big(x+\frac{y}{2},t\Big) + Z\Big(x-\frac{y}{2},t\Big) \Big\}\, dy \nonumber \\
& \qquad \qquad \times W_x(x,k-\lambda,t) \, d\lambda , 
\end{align}
where we have introduced the auxiliary function
\begin{equation*}
 Z(x,t)= \frac{1}{2\pi}\int_{\R} W(x,k,t)\,dk,
\end{equation*}
which represents the (longitudinal) density energy. The expression (\ref{eq:HirotaWigner}) may be readily obtained from  Eq.(\ref{eq:HE}) by applying the transformation (\ref{eq:WignerTransf}) (see \ref{App1} for further details). Accordingly, the wave amplitude is given by \cite{segur20051}
\begin{equation}\label{eq:waveamp}
E(t)=\frac{1}{2\pi}\int_{\R^{2}} W(x,k,t)\,dkdx,
\end{equation}
which will be conserved for all parameter combinations in any given solution of the ordinary Hirota equation in absence of gain/loss mechanism \cite{achilleos20151}. The latter will produce an amplification or attenuation of the wave amplitude that shall be fully treated in Sec.\ref{Sec:LD} (see the Eq.(\ref{EELD})).

Basically, the MI analysis of the wave spectra consists of introducing a perturbation of small strength $\varepsilon\ll 1$ around a stationary distribution $W_{0}$ 
\cite{
picozzi20141,
hall20021,
helczynski20031,
onorato20031}, 
i.e.  
\begin{equation}
W(x,k,t)=W_0(k) + \varepsilon \,F_{K,\Omega}(k) \, e^{i(Kx - \Omega t)},
\label{eq:ansatz}
\end{equation}
with $K\in \mathbb{R}$ and $\Omega  \in \mathbb{C}$ being the modulation wave number and modulation frequency, respectively. The term $F_{K,\Omega}(k)$ takes into account the spatial profile of the perturbation \cite{shukla20051}. This plays no significant role in the MI analysis at leading order
\cite{
shukla20051,
athanassoulis20171,
onorato20031} (see the dispersion relation \eqref{eq:dispersion relation} below), and it is assumed to be an integrable function for mathematical convenience. It is widely known that the Benjamin-Feir instability corresponds to the case in which  $W_{0}(k)$ is the Wigner function of a plane wave solution, i.e.
\begin{equation}
\psi_{pw}=\frac{\psi_0}{\sqrt{2\pi}} e^{i  \theta\frac{ \psi_0^2}{2\pi}t}.
\label{PWS}    
\end{equation}
In this work we shall refer to incoherent MI when $W_{0}(k)$ is the Wigner function of a plane wave with a complex amplitude modeled by an stochastically varying phase term $\varphi(x)$ 
\cite{
shukla20051,
hall20021,
anderson20041}, i.e. $\psi=\psi_{pw}e^{i\varphi(x)}$. This essentially represents an additional non-monochromatic contribution to the stationary plane-wave solution that accounts for the spatial incoherence of light pulses in optical fibers \cite{soljanic20001, hall20021}. In this sense, we could understand the Benjamin-Feir instability as the "coherent" side of the MI. It is important to realize that, though $\psi$ is not an exact solution of the Hirota equation, it represents a stationary solution at leading order of the perturbative analysis (\ref{eq:ansatz}) in the Wigner-function framework upon imposing appropriate conditions (see the dispersion relation (\ref{eq:dispersion relation}) below).

Clearly, if the imaginary part of $\Omega$, denoted by $\text{Im}\Omega$, takes positive values, it immediately follows that the perturbative part blows up at a finite time (such as $\text{exp}(\text{Im}\Omega \,t)$), and thus, the MI would dominate the long-time dynamics. This is commonly referred to as the MI growth rate or MI gain associated to the stationary solution 
\cite{
chen20171,
ankiewicz20101}.
In this context, we define the unstable spectrum as composed of all modes $K_{I}$ giving rise to a growing wave amplitude for some $t>0$, i.e. $\text{Im}\Omega>0$ \cite{segur20051,janssen20031}.
Importantly, in the case that this spectrum contains the zero wave-number $K=0$ (or sufficiently low modulation wave number) as a limiting case, this instability is referred to as MI baseband \cite{chen20171,chen20151}. As stressed in the introduction, it is broadly accepted that this type of instability is a prerequisite for the formation of rogue waves in generalized forms of the NLSE
\cite{
baronio20151,
baronio20141,
chen20151} 
or Hirota equation \cite{ankiewicz20101,wang20161,wen20181}. Based on this conjecture, in Sec. \ref{EOEWE} we discuss the emergence of extreme wave events in feasible experimental scenarios subjected to the linear and Lorentzian spectrum damping effects.

In a first approach, the unstable spectrum is characterized from the dispersion relation for the perturbative frequency obtained from linearizing the equation (\ref{eq:HirotaWigner}) around the stationary solution 
\cite{
fedele20001,
hall20021,
onorato20031}. Concretely, after a linear perturbative analysis of the Eq.(\ref{eq:HirotaWigner}) once substituted the ansatz (\ref{eq:ansatz}) (further details can be found in the \ref{App2}), one arrives to the leading expression providing the desired (nonlinear) dispersion relation for a given stationary distribution,
\begin{eqnarray}
1&+&\frac{1}{4\pi}
\int_{\R}\Big\{\Omega
+\frac{\gamma}{4} K^3
- \frac{\alpha \psi_0^2}{2\pi}K
+\big(- 2\beta k + 3\gamma  k^2 \big) \, K\Big\}^{-1} \nonumber \\ 
&\times& \Big\{ (\alpha k - \theta)\big(W_0(k+K/2)-W_0(k-K/2)\big) \nonumber  \\
&+&\alpha \frac{K}{2}\big(W_0(k+K/2)+W_0(k-K/2)\big)\Big\} dk=0,
\label{eq:dispersion relation}
\end{eqnarray}
with
$$\psi_0^2= \int_\R W_0(k) \, dk.$$\\
which reduces to the well-known result for the NSLE in the limit $\alpha=\gamma=0$ and $\beta=1$ \cite{hall20021,onorato20031}. Now upon replacing a stationary distribution, we shall obtain the MI gain by computing the above integral by means of the standard contour integration techniques (the interested reader can find further details in \ref{App2}). Here we would like to notice that an identical linearization procedure in the case of a non-vanishing attenuation/amplification coefficient (i.e. $\eta(t) \neq 0$) can be followed independently of the choice of the stationary distribution $W_{0}(k)$. This is due to the fact that the gain/loss Hirota equation (\ref{eq:HE}) can be cast in the form of a time-dependent Hirota equation (\ref{eq: NewMHE}) (without linear gain/loss term) as explained in further detail in Sec.\ref{Sec:LD}: the latter equation admits a stationary plane-wave-like solution (\ref{PWSDIS}) that returns the ordinary plane wave (\ref{PWS}) for a vanishing gain/loss coefficient $\eta(t) = 0$.

Compared to the NLSE case, the dispersion relation (\ref{eq:dispersion relation}) has an additional pole which exhibits an intricate relation with $\Omega$ when $\gamma \neq 0$ (see the discussion around the Eqs. from (\ref{Aroots0}) to (\ref{Aroots2}) in the \ref{App2}). This makes difficult to analytically elucidate the MI gain and the associated unstable spectrum for the broad set of parameter values, so we shall perform both analytical as well as numerical analysis for distinct values of the TOD strength. Without loss of generality, we shall assume that $\beta$, $\alpha$ and $\gamma$ are positive \cite{chen20171}. Additionally, we shall take $\theta>0$, since the MI appears for the self-focusing case as it is well-known from previous studies in the Hirota equation \cite{ankiewicz20101}. Here it is important to realize that the meaning of both modulation parameters $\Omega$ and $K$ depends on the analyzed experimental platform \cite{chen20131}: while for surface gravity waves $\Omega$ and $K$ play the role of the frequency and wave number of the unstable mode \cite{zhang20171,zhang20191}, their roles are interchanged when studying ultrashort pulses \cite{ankiewicz20101}.

\subsection{Benjamin-Feir instability}\label{Sec:BF}

Before addressing the incoherent modulation instability, we briefly study the standard case of $W_0$ representing a plane wave
in order to show that our treatment reproduce previous well-known results from the Hirota equation. According to the standard Benjamin-Feir instability analysis \cite{zakharov20091}, we must take the Wigner transform associated to the plane wave solution \cite{ankiewicz20101,achilleos20151},
\begin{equation}
W_0(k)= \psi_0^2 \delta(k),
\label{eq: deltaWigner}
\end{equation}
which inserted in (\ref{eq:dispersion relation}) returns the following dispersion relation after some straightforward manipulation 
(see \ref{App2}1),
\begin{equation}\label{eq:growthrate}
\Omega = 
\frac{\alpha \psi_0^2 }{2\pi} K - \gamma K^3
\pm i\beta K^2 
\sqrt{\frac{\theta \psi_0^2}{2\pi\beta K^2} - 1} .
\end{equation}
This result is in complete agreement with previous works based on the wave function approach 
\cite{
ankiewicz20101,
guo20161, 
chabchoub20151,
achilleos20151}.
From here, it is immediate to obtain the largest MI gain,
\begin{equation*}
\text{Im}\Omega_{max}
=
\frac{\theta\psi^{2}_{0}}{4\pi},
\end{equation*}
which occurs at 
$K_{max}
=
\sqrt{\theta\psi_0^2/4\pi\beta}$. 
Additionally, from (\ref{eq:growthrate}) we compute the unstable spectrum of the wave number which may eventually give rise to MI, i.e.
\begin{equation}
K_{I}\in \bigg[-\sqrt{\frac{\theta\psi_0^2}{2\pi\beta}}, \sqrt{\frac{\theta\psi_0^2}{2\pi\beta}}\bigg],
\label{IKS}
\end{equation}
which manifests that the MI for the Hirota equation is baseband type for any $\theta>0$ \cite{chen20171}. Interestingly, the Benjamin-Feir instability analysis indicates that the MI is solely produced by the Kerr nonlinearity. Contrary to what one could expect, the SS interaction plays no role in the emergence of instability, 
neither the TOD, in agreement with the discussion in 
\cite{
ankiewicz20101,
zhang20091}. 
In other words, the Benjamin-Feir instability is insensitive to the higher-order nonlinear effects, and as a result, it could ignore the emergence of extreme waves situations in which the nonlinear stage is dominated by these instead of the Kerr interaction.

\begin{figure*}[]
\includegraphics[scale=0.29]{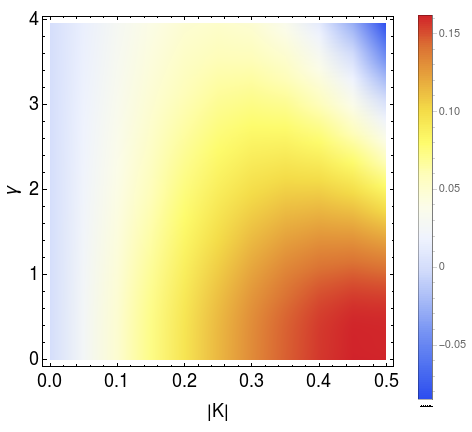}
\includegraphics[scale=0.359]{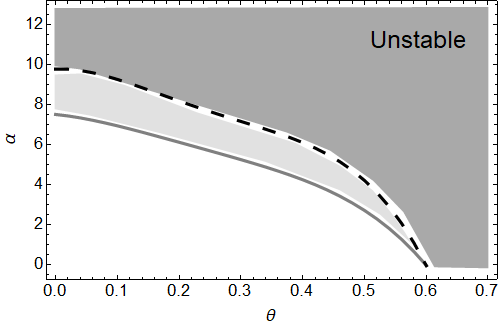}
\includegraphics[scale=0.435]{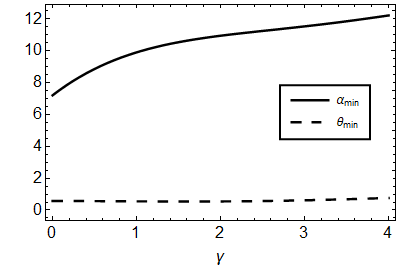}
\caption{(color online). (Left) Density plot of the MI gain as a function of the modulation wave number and TOD strength for fixed values of $p_{0}=0.15$, $\theta=4$ and $\alpha=2$. (Central) Contour plot of the instability region as a function of the Kerr $\theta$ and the SS $\alpha$ terms for two values of the TOD strength: the lightgray solid and black dashed lines correspond to $\gamma=0$, and $\gamma=1$, respectively. Similarly, the lightgray and darkgray shadowed areas indicate the unstable regions obtained for the corresponding values of $\gamma$ (notice they overlap beyond the black-dashed line). The wave number was taken close to the machine numerical zero $K=-10^{-8}$. (Right) The minimum values $\alpha_{min}$ and $\theta_{min}$ as functions of TOD contribution. In all the figures the problem parameters are rephrased to the dimension scales in which $\beta=1$, $p_{0}=0.15$ and $\psi_{0}=1$.\label{Fig1}} 
\end{figure*}

\subsection{Incoherent modulation instability}\label{Sec:IMI}

Let us turn the attention to the study of the MI in presence of Lorentzian spectrum damping effects. As stated in the Introduction, this is accomplished by following the incoherent modulation instability analysis 
\cite{
picozzi20141,
chabchoub20151,
lumer20161,
guasoni20171}: now we address the modulation instability due to the stationary distribution $W_{0}(k)$ retrieved by $\psi= \psi_{pw}e^{ i\varphi(x)}$, with $\varphi(x)$ being a randomly varying phase. This implies that, once replaced $\psi$ in the definition of the Wigner function (\ref{eq:WignerTransf}), we must perform the ensemble average of the phase before carrying out the integral in the variable $y$ \cite{hall20021} (this average shall be indicated by $\left\langle \cdots \right\rangle $). Since $\varphi(x)$ arises from an environmental noise destroying the coherence length of the propagating pulse in the vast majority of practical situations 
\cite{
shukla20051, 
marklund20061,
anderson20041,
soljanic20001}, we may consider this to yield an exponentially decaying amplitude effect, i.e.   
\begin{equation}
\left\langle \psi(x+y/2,t)\psi^{*}(x-y/2,t)\right\rangle ={\psi_{0}}^{2}e^{-p_{0}|y|},
\label{SSANT}
\end{equation}
where $p_0 >0$ can be thought of as a spatial coherence length \cite{soljanic20001} or frequency bandwidth \cite{kibler20121} depending on the experimental platform of interest. Notice that this choice satisfies the expected condition of translational invariance of the leading-order solution (\ref{eq:ansatz}). Indeed the choice (\ref{SSANT}) characterizes the common situation in the context of ultrashort pulses propagating in optical fibers 
\cite{
shukla20051,
marklund20061,
hall20021,
hall20041,
helczynski20031,
anderson20041,
soljanic20001} 
as well as surface gravity waves \cite{onorato20031,onorato20061}. As stated in the introduction, we shall follow the convention of previous treatments 
\cite{
picozzi20141,
shukla20051,
hall20021} 
and reefer to this as the Lorentzian spectrum damping: $p_{0}$ represents the strength of the effective Lorentzian spectrum damping for electromagnetic waves propagating in nonlinear media
\cite{
fedele20001,
fedele20002}. 
The corresponding Wigner transform takes a Lorentzian shape 
\cite{
hall20021,
anderson20041,
soljanic20001}, i.e.
\begin{equation}
W_0(k)= \frac{\psi_0^2}{\pi} \frac{p_0}{k^2 + p_0^2}, 
\label{LW0}  
\end{equation}
which boils down into the previous plane-wave solution (\ref{eq: deltaWigner}) in the strict limit $p_0\rightarrow 0$, retrieving the standard Benjamin-Feir analysis
\cite{
ankiewicz20101,
guo20161,
soljanic20001,
onorato20031}. Let us notice that the stationary distribution obtained from the so-called JONSWAP spectrum, which describes the spectrum of the free surface elevation of ocean waves in several meteorological conditions, is well approximated by the Lorentzian shape (\ref{LW0}) in the limit of a narrow bandwidth \cite{onorato20031}. Interestingly, this limit has been experimentally explored in order to observe the formation of rogue waves in partially coherent waves in the context of the NLSE \cite{elkoussaifi20181}.

Going back to the dispersion relation (\ref{eq:dispersion relation}), 
one may see that the aforementioned additional pole simplifies by taking a vanishing TOD $\gamma=0$ (see the Eqs. (\ref{Aroots1}) and (\ref{Aroots2}) in \ref{App2}), which substantially facilitates the MI analysis. For the clarity of exposition, we first focus on the situation where the TOD effect is sufficiently small compared to other nonlinearity interactions, and it can be negligible. In an optical scenario, this could be regarded as the case in which the pulse wavelength lies outside the dispersion regime of the fiber 
\cite{
kibler20151,
conforti20151}.

\subsubsection{Incoherent modulation instability for vanishing TOD effects}\label{Sec:IMI1}

By following a similar procedure as to obtain the Eq.(\ref{eq:growthrate}), we compute the dispersion relation (see the details in \ref{App2}2)
\begin{equation}
\Omega = \frac{\alpha\psi_0^2 }{2\pi} K 
+ 2ip_0\beta K \pm i\beta K^2 \, \sqrt{\frac{\psi_0^2}{2\pi\beta K^2} (\theta - ip_0 \alpha) - 1},
\label{Omega Lorentz}
\end{equation}
which returns the well-known result of the NLSE when the self-steppening effects are negligible (i.e. $\alpha=0$) \cite{hall20021,onorato20031}. At first sight, we may appreciate that the instability growth rate significantly differs from the result (\ref{eq:growthrate}) obtained via the Benjamin-Feir analysis. As similarly noted in the context of four-state atomic systems \cite{shukla20051,marklund20061}, the Eq. (\ref{Omega Lorentz}) unveils an intricate interplay between the higher-nonlinear corrections and a frequency bandwidth: the contribution from the SS interaction to the instability growth rate exclusively couples to the Lorentzian spectrum damping. In other words, though large values of $p_{0}$ may completely suppress the (first-order) MI, a moderate Lorentzian spectrum damping assists the instability effects owing to the higher-order nonlinear interaction (recall that this is absent in the Benjamin-Feir instability growth rate). This directly suggests that the Lorentzian spectrum damping may prove beneficial for the rising of higher-order nonlinearities. 

After some readily algebra in Eq.(\ref{Omega Lorentz}) (see the \ref{App2}3 for a detail derivation), one also finds the unstable spectrum
\begin{align} \label{Eq:SPC} 
&K_{I}
\in 
\Bigg[ -\sqrt{ \frac{\psi_0^2 \theta}{2\pi\beta}+ \bigg(\frac{\psi_0^2 \alpha}{8\pi\beta}\bigg)^2-4p_0^2}, 
 \sqrt{ \frac{\psi_0^2 \theta}{2\pi\beta}+ \bigg(\frac{\psi_0^2\alpha}{8\pi\beta}\bigg)^2 -4p_0^2}\Bigg].  
\end{align}
From the interval (\ref{Eq:SPC}) it is clear  that there exists a competition between the Lorentzian spectrum damping and the nonlinear interactions
\cite{
soljanic20001,
fedele20001,
onorato20031,
lumer20161}.
Furthermore, the interval (\ref{Eq:SPC}) reveals that the MI is of baseband type as well, for any value of the Lorentzian spectrum damping below the modulational instability threshold
\begin{equation}
4 p_{0}< \frac{\psi_0^2 \theta}{2\pi\beta}+ \bigg(\frac{\psi_0^2 \alpha}{8\pi\beta}\bigg)^2.
\label{LDD}    
\end{equation}
Otherwise the Lorentzian spectrum damping completely suppresses the (first-order) instability. Unfortunately, we are still lacking of a full physical picture of the mechanism behind the Lorentzian spectrum damping in either a nonlinear optical medium 
\cite{
fedele20001,
hall20021} 
or surface gravity waves \cite{onorato20031} partially because there is no rigorous equivalence between the Landau damping from plasma physics and the one studied here \cite{fedele20002}.

\subsubsection{Incoherent modulation instability for non-vanishing TOD effects}\label{Sec:IMI2}

Now we discuss the action of the TOD effects upon the MI. From the Eq.(\ref{eq:growthrate}) it follows that the growth rate characteristic of the Benjamin-Feir instability is insensitive to this. However, this situation drastically changes in presence of the Lorentzian spectrum damping \cite{helczynski20031}, and it is less clear how the TOD influences the instability structure owing to higher-nonlinear interactions \cite{shukla20051,marklund20061}. We illustrate in the Figure \ref{Fig1} the results obtained from the numerical computation of the (maximum) MI gain (i.e. $\text{Im}\Omega_{max}$) with $\gamma \neq 0$. The left panel depicts the MI gain as a function of the modulation wave number and TOD strength. A quick glance reveals that the MI is a baseband type as well for any given $\gamma$, as the MI gain softly decreases as $|K|$ gradually goes to zero \cite{chen20171}. For a fixed value of the later, we may also appreciate that the MI decays for higher $\gamma$, which is in accordance with the intuition that the TOD is detrimental for the emergence of instability structure, it plays the role of a stabilizing agent. 

The central panel of Figure \ref{Fig1} represents the unstable region (see the lightgray and darkgray shadowed areas) associated to the modulation wave numbers close to zero ($|K|\sim 10^{-8}$), as a function of the Kerr and the SS terms. Accordingly, the gray solid line, which is obtained for $\gamma=0$, is determined from the inequality (\ref{LDD}). Observe that the unstable region shrinks as the TOD strength increases (the darkgray shadowed area corresponds to a non-vanishing $\gamma$) in complete agreement with the above discussion. Interestingly, this plot also manifests that the TOD substantially suppresses the MI effects owing to the SS nonlinearity. That is, for $\theta=0$ we may observe that the minimum value of the SS strength giving rise to MI, denoted by $\alpha_{min}$, is demanded to grow with $\gamma$. Conversely, the minimum value due to the Kerr nonlinearity, i.e. $\theta_{min}\approx 0.3$, is barely affected by the stabilizing effect due to the TOD.

This feature is better illustrated in the right panel of Figure \ref{Fig1}, which depicts the behavior of $\theta_{min}$ and $\alpha_{min}$ in terms of $\gamma$. While $\theta_{min}$ seems to remain almost constant, the $\alpha_{min}$ substantially grows by an algebraic increment of, at least, two orders of magnitude in comparison with $\theta_{min}$. This directly shows that the instability structure characteristic of the Kerr nonlinearity is significantly robust to the TOD. On the other side, though it is not shown here, we find out that the Lorentzian spectrum damping may eventually cancel the MI gain, and thus, suppress the emergence of instability due to the Kerr as well as higher-order nonlinearities, as expected from the subsidiary condition (\ref{LDD}).

\section{Wigner instability analysis with linear gain/loss effects}\label{Sec:LD}

\begin{table*}
\begin{center}
\begin{tabular}{c|c|c|c|c|c|c}
\hline
\hline
\quad \quad Experimental platform \quad \quad  &\quad \quad $|\beta|$ \quad \quad & \quad \quad $|\theta|$ \quad \quad &  \quad \quad  $|\alpha|$ \quad \quad & \quad \quad $|\gamma|$ \quad \quad & \quad \quad $p_{0}$\quad \quad  & \quad \quad $|K_{I} |$ \quad \quad \\
\hline
Ultrashort pulses in optical fibers           &  0.005-88      & 0.01-10     &  $ 6-7\times 10^{-3}$ & 0.01-0.04 & 0.089-0.139 & $\lesssim 0.61$  \\ 
(Refs.
\cite{kibler20101,
simaeys20011,
yan20131,
kraych20191,
elkoussaifi20181,
tikan20171,
solli20121,erkintalo20121})    &   ($ps^2/km$)    &  ($1/W km$)  & ($ps/W km$) & ($ps^3/km$) & (THz) & (THz)  \\ \hline
Surface gravity waves in deep water     & 0.01-12.5        &   0.001-729    & 0.1-80 & 0.1-4 & 0.06-0.1 &  $\lesssim 7$ \\ 
    (Refs.\cite{chabchoub20111,
peric20151,
kimmoun20161,
kimmoun20171,
onorato20031,
onorato20061,
elkoussaifi20181,
onorato20051,
fedele20161,
chabchoub20121})        &  ($m^2/s$)   & ($1/s m^{2}$)   &  ($1/s m$) & ($m^3/s$) & ($1/m$) & ($1/m$)\\ \hline
\hline
\end{tabular}
\end{center}
\caption{The unstable spectrum for feasible experimental scenarios when we disregard the linear amplification/attenuation coefficient (i.e. $\eta(t)\rightarrow0$). The seventh column shows the modulation wave number (numerically) computed for the displayed values (expressed in SI units) of the problem parameters. These values are indicative and were extracted from previous theoretical and experimental studies in the realm of the NLSE and Hirota equation. Notice that the variable $K_{I}$ plays the role of a modulation frequency in the MI analysis of optical fiber experiments.}
\label{tab:table1}
\end{table*}

In this section we address the emergence of instability in the presence of a non-vanishing linear gain/loss contribution 
$i\eta(t)\psi$. This reproduces the laboratory conditions in a broad range of experimental situations: for instance,  
$\eta(t)<0$ regards the dissipation rate due to the friction with the walls in deep-water wave  
\cite{
kimmoun20161,
segur20051,
kimmoun20171,
onorato20121} 
as well as photon losses in light propagation experiments
\cite{
hao20041,
yan20131,
kraych20191,
sagnes20101}, 
or alternatively, $\eta(t)>0$ represents a wind forcing effect \cite{onorato20121} as well as a spontaneous emission noise generated by optical amplifiers in fibers \cite{xu20031,lucas20171}. 

Accordingly, after doing an appropriate change of variables \cite{segur20051,onorato20121}, i.e.
\begin{equation}
    \psi(x,t)\rightarrow \phi (x,t) e^{\int_{0}^{t}\eta(\tau) d\tau},
    \label{CHHD}
\end{equation}
the Hirota equation endowed with a linear absorption/amplification coefficient can be cast in the usual form, i.e.
\begin{equation}
    i\phi _t + i\alpha (t) |\phi |^{2} \phi _x + \beta \phi _{xx} + i\gamma \phi _{xxx} + \theta (t) |\phi |^2 \phi  = 0,
\label{eq: NewMHE}    
\end{equation}
with $\alpha (t) = \alpha e^{2\int_{0}^{t}\eta(\tau)d\tau}$ and $\theta (t) = \theta  e^{2\int_{0}^{t}\eta(\tau)d\tau}$. Importantly, the Eq.(\ref{eq: NewMHE}) also admits a stationary plane-wave-like solution, i.e.
\begin{equation}
\phi_{pw}=\frac{\phi_{0}}{\sqrt{2\pi}}\text{exp}\Bigg(i\frac{\theta\phi_{0}^2}{2\pi}\int_{0}^{t}e^{2\int_{0}^{\tau}\eta(\tau')d\tau'}d\tau \Bigg).
\label{PWSDIS}
\end{equation}
Additionally, the linear gain/loss contribution introduces a time-dependent coefficient in front of the wave amplitude \eqref{eq:waveamp}, i.e.
\begin{equation}
E_{\psi}(t)= E_{\phi}(t)  e^{2\int_{0}^{t}\eta(\tau)d\tau}  .
\label{EELD}    
\end{equation} 
While the (linear) gain intensifies the amplitude of small instabilities, the (linear) loss can completely suppress the wave amplitude in the short time unless instability comes into play \cite{segur20051}. By appealing an exponentially exploding perturbative solution once substituted (\ref{CHHD}) in (\ref{eq:ansatz}), and after some manipulation we arrive to the subsidiary condition
\begin{equation}
- \frac{2}{t}\int_{0}^{t}\eta(\tau)d\tau<\text{Im}\Omega_{max}, \ \ t>0,
\label{SSDC}
\end{equation}
which determines the emergence of the instability under linear dissipation in the first stages of the dynamics (e.g. the inequality (\ref{SSDC}) is akin to the condition (\ref{LDD})). Since the structure of Eq.(\ref{eq: NewMHE}) is formally equivalent to (\ref{eq:HE}), we may carry out a similar procedure as before by replacing $\psi_{pw}\rightarrow \phi_{pw}$ in the study of both the Benjamin-Feir and incoherent instabilities. From the Wigner transform definition (\ref{eq:WignerTransf}), it is readily to see that such replacement returns stationary solutions equal to the Eqs. (\ref{eq: deltaWigner}) and (\ref{LW0}), and thus, the MI analysis in the Wigner framework retrieves identical results for the dispersion relation (see the Eqs.(\ref{eq:growthrate}) and (\ref{Omega Lorentz})) as well as unstable spectrum (see the Eqs.(\ref{IKS}) and  (\ref{Eq:SPC})) once substituted the coefficients $\alpha\rightarrow \alpha (t)$ and $\theta\rightarrow\theta(t)$. For instance, for the simple case of a time-independent coefficient, say $\eta(t)=\eta_{0}$, and vanishing Lorentzian spectrum damping (i.e. $p_{0}=0$), the condition (\ref{SSDC}) directly yields
\begin{equation}
 K^4-\frac{\psi_{0}^{2}\theta(t) }{2\pi\beta}e^{2\eta_{0} t}K^2+\frac{4 \eta_{0}^2}{\beta^2} <0.  
\label{SSDCTI}
\end{equation}

Observe that the linear gain/loss influences the unstable spectrum as well: this is effectively increased or reduced by the amplification/attenuation coefficient. As a consequence, the linear amplification can substantially diminish the degrading effects owing to the Lorentzian spectrum damping. On the other side, the (linear) dissipation will eventually stabilize any unstable stationary solution despite how small is the coefficient $\eta$ \cite{segur20051,onorato20121} (e.g. see the inequality (\ref{SSDCTI}) when $t\rightarrow \infty$). In other words, we may expect that (linear) dissipation inhibits the appearance of extreme wave situations in the long time limit, nonetheless this may survive at short time scales $t\ll|\eta^{-1}|$ provided the subsidiary condition (\ref{SSDC}) holds.\\

To summarize, we have shown that the incoherent modulation analysis reveals new features about the MI of the Hirota equation which are unappreciated by following the standard Benjamin-Feir treatment \cite{zakharov20091}: we found out that the Lorentzian spectrum damping in conjunction with the SS interaction enhances up the instability features rendering an extended unstable spectrum \cite{shukla20051,marklund20061}. On the other side, the instability structure arising from the Kerr interaction is significantly resilient to the stabilizing effects of the TOD, while the SS contribution may be eventually cancelled out. We finally saw that (linear) gain/loss effects dominate the instability structure for moderates values of the nonlinear interactions.

\section{Extreme wave events}\label{EOEWE}

Relying on our previous results, we now discuss in parallel the attainable observation of remarkable wave events (e.g. FPUT recurrences or Peregrine solitons) in two specific experimental situations: ultrashort pulses traveling in optical fibers
\cite{
kibler20101,
simaeys20011,
yan20131,
kraych20191,
elkoussaifi20181,
tikan20171,
solli20121,
erkintalo20121}, 
and surface gravity waves in deep water
\cite{
chabchoub20111,
peric20151,
kimmoun20161,
kimmoun20171,
onorato20031,
onorato20061,
elkoussaifi20181,
onorato20051,
fedele20161,
chabchoub20121}. 
As previously mentioned, it is widely recognized that the formation of rogue waves may occur whenever the relevant system exhibits a
baseband MI \cite{chen20171,wang20161,wen20181}. Concretely, it was shown for generalized NLSE that the Peregrine soliton solution is closely related to the existence of baseband MI \cite{baronio20151,baronio20141,chen20151}. Similarly, the so-called Akmediev breathers, which are fundamental for the appearance of the rogue waves in the NLSE \cite{ankiewicz20141} and the Hirota equation 
\cite{
ankiewicz20101,
tao20121}, 
are direct consequence of the MI dynamics 
\cite{
dudley20141,
dudley20091}. 
Based on these previous theoretical findings and our current results, we could conclude that the observation of extreme wave events arising from MI conditions are accessible within the available lab technologies despite the unavoidable degrading effects owing to the linear or Landau-like dampings. 

Let us first ignore the linear gain/loss effects. Table \ref{tab:table1} displays the instability spectrum followed from our treatment for typical experimental values of the problem parameters in both platforms. Our analysis indicates that the MI characteristic of the Hirota equation is fairly robust to realistic Lorentzian spectrum damping effects, and signatures of the unstable dynamics could manifest for a broad range of feasible modulation frequencies $\lesssim 0.61$THz in optical fiber scenarios, or alternatively, modulation wave numbers $\lesssim 7m^{-1}$ in the case of surface gravity waves experiments.  Interestingly, these results are in quantitative agreement with previous experimental realizations in the realm of the NLSE, for which the damping effects were disregarded. More specifically, in Refs. \cite{simaeys20011,kibler20151} it was demonstrated the generation of the FPUT recurrence and Peregrine solitons in optical fibers by Benjamin-Feir MI with modulation frequencies 340-520GHz and 196-278GHz, repectively. Similarly, in Ref. \cite{chabchoub20111} it was experimentally probed the formation of Peregrine solitons in water wave tanks for a carrier wavenumber around $11.63m^{-1}$. 

Let us first ignore the linear gain/loss effects. Table \ref{tab:table1} displays the instability spectrum followed from our treatment for typical experimental values of the problem parameters in both platforms. Our analysis indicates that the MI characteristic of the Hirota equation is fairly robust to realistic Lorentzian spectrum damping effects, and signatures of the unstable dynamics could manifest for a broad range of feasible modulation frequencies $\lesssim 0.61$THz in optical fiber scenarios, or alternatively, modulation wave numbers $\lesssim 7\text{m}^{-1}$ in the case of surface gravity waves experiments.  Interestingly, these results are in quantitative agreement with previous experimental realizations in the realm of the NLSE, for which the damping effects were disregarded. More specifically, in Refs. \cite{simaeys20011,kibler20151} it was demonstrated the generation of the FPUT recurrence and Peregrine solitons in optical fibers by Benjamin-Feir MI with modulation frequencies 340-520GHz and 196-278GHz, repectively. Similarly, in Ref. \cite{chabchoub20111} it was experimentally probed the formation of Peregrine solitons in water wave tanks for a carrier wavenumber around $11.63\text{m}^{-1}$. 

As illustrated in Sec.\ref{Sec:LD}, the effects of the linear dissipation is twofold: to attenuate the instability wave amplitude as well as to shrink the unstable spectrum. To analyse the first degrading effect let us focus on the subsidiary condition (\ref{SSDC}) for a time-independent dissipation $\eta(t)=\eta_{0}$ and vanishing TOD (recall the MI gain is then determined by the Eq.(\ref{Omega Lorentz}) once substituted $\alpha\rightarrow \alpha (t)$ and $\theta\rightarrow\theta(t)$). By replacing certain characteristic values of the problem parameters and the modulation number $K_{I}$ from the table \ref{tab:table1}, we may obtain a rough estimation for the damping rate beyond which the dissipation dominates the wave amplitude dynamics: $\eta_{0}<0.5\times 10^{-3}\text{km}^{-1}$ for ultrashort pulses, and $\eta_{0}<0.49\text{s}^{-1}$ for surface gravity waves. A direct comparison to the typical loss in standard optical fibers (for instance, $\simeq0.18$ dB/km \cite{kimmoun20171,erkintalo20121}, i.e. $2\%$ of losses in 500m) manifests that the attenuation amplitude does not represent a crucial barrier for the observation of MI emerging from the Hirota equation in the early stage. On the other side, the harmful effects upon the unstable spectrum (given by the Eq.(\ref{Eq:SPC})) can be expressed as follows in the short time scale $t\ll \eta_{0}^{-1}$,
\begin{equation}
|K_{I}|^2=   |K_{I}|_{\eta_{0}=0}^2+2\eta_{0} t\Bigg(\frac{\psi_0^2 \theta}{2\pi\beta}+ \bigg(\frac{\psi_0^2 \alpha}{8\pi\beta}\bigg)^2\Bigg)+\mathcal{O}\Big((\eta_{0}t)^2\Big), \label{KLDI}
\end{equation}
where $|K_{I}|_{\eta_{0}=0}$ corresponds to the values given by Table \ref{tab:table1}. Accordingly to the result (\ref{KLDI}), we must get a compromise between the strength of the nonlinear interactions and the dissipative coefficient, otherwise the unstable spectrum will decrease relatively fast making hard to match the MI conditions in the first stages of dynamics. For instance, in experimental realizations of the FPUT recurrence in water tanks \cite{kimmoun20161} (see also \cite{kimmoun20171}), a realistic dissipative coefficient $ \eta_{0}$ was estimated to vary from $1.3 \times10^{-3}\text{m}^{-1}$ to $1.6\times10^{-3}\text{m}^{-1}$ (notice that $K_{I}$ represents a modulation frequency in this case). This leads to the unstable spectrum to decay with a ratio $\sim 6.5(\text{m}s^{2})^{-1}$, which is within the typical spatial and time scales in water tank experiments. Additionally, it has been recently shown both analytically and numerically that periodic anomalous wave events in the NLSE may be observed provided the attenuation or amplification parameter is small in comparison with certain characteristic time $T$ \cite{coppini20191}. Extrapolating this condition to our more general situation, we obtain
\begin{equation}
\eta_{0} T^2\Bigg(\frac{\psi_0^2 \theta}{2\pi\beta}+ \bigg(\frac{\psi_0^2\alpha}{8\pi\beta}\bigg)^2\Bigg)\ll 1, \label{CHT}  
\end{equation}
Clearly, the subsidiary condition (\ref{CHT}) manifest that the ratio between the Kerr interaction and velocity dispersion dominates the aforementioned characteristic time, and thus, the attainability of periodic extreme wave events.

\section{Concluding remarks}\label{Sec:CR}

In this work we have addressed the unstable structure of the damped Hirota equation by following a (first-order) incoherent MI analysis in the Wigner-function framework. Our results stress out that this incoherent analysis renders a more complete description of the instability in presence of significant higher-order contributions than the standard Benjamin-Feir approach. While the latter treatment completely disregards the instability effects owing to the higher-order nonlinearity, the former analysis unveils an intriguing interplay between this and the Lorentzian spectrum damping: contrary to most common intuition from the NLSE, a moderate Lorentzian spectrum damping may assist the MI arising from the SS interaction. Unlike previous works, we also show that the stabilizing effects due to the TOD barely influences the unstable structure owing to the Kerr interaction in comparison with the SS nonlinearity. Additionally our treatment proves convenient to study the impact of (linear) time-dependent gain/loss terms, we show that these may eventually dominate the emergence of the MI independent of their strength. Beside, we contrast our results with current experimental platforms (e.g ultrashort pulses in optical fibers and surface gravity waves in deep waters), and identify the regimes where extreme wave events driven by the MI may appear despite the degrading effects due to the Lorentzian spectrum and linear dissipative dampings.

Remarkably, the recent experimental progresses in the observation of FPUT recurrence in dissipative water tanks 
\cite{
kimmoun20161,
kimmoun20171}, 
the generation of breathing solitons in amplified/attenuated microresonators \cite{lucas20171} or in noise-driven optical fibers \cite{kraych20191}, as well as the creation of optical solitons in non-ideal photonic crystal waveguides \cite{sagnes20101} open new avenues to get a deeper understanding of the formation of extreme wave events in more realistic situations. In this sense, the present treatment could provide a valuable theoretical support to plan a new series of experiments to study interesting wave events driven by the MI dynamics when the linear dissipative and Lorentzian spectrum damping effects play a prominent role, as occur for ultrashort pulses or surface gravity waves in less controlled mediums.

\section*{Acknowledgments}
The authors warmly thanks J.M. Soto-Crespo for useful discussions. This work is supported by the Nazarbayev University Faculty Development Competitive Research Grants Program, grant number 110119FD4544. This material is also based upon work supported by the Air Force Office of Scientific Research under award number FA2386-18-1-4019.

\appendix

\section{Wigner-Hirota equation derivation}\label{App1}

This section is devoted to illustrate the derivation of the Eq.(\ref{eq:HirotaWigner}) by starting from expression (\ref{eq:HE}) once performed the Wigner-Moyal transform given by (\ref{eq:WignerTransf}). For seek of clarity it is convenient to rephrase the latter in terms of arbitrary complex functions $u,v$, i.e. \cite{athanassoulis20171,athanassoulis20201}
\begin{equation}
W[u,v] = \int_{\R} e^{-iky} u(x+\frac{y}{2},t) \, v^{*} (x-\frac{y}{2},t) dy,
\label{WMT2}
\end{equation}
so that the Wigner-Moyal transform (\ref{eq:WignerTransf}) is retrieved by $W=W[\psi,\psi]$. Following the definition (\ref{WMT2}), we may rewrite the Hirota equation (\ref{eq:HE}) as follows  
\begin{align*}
 W_t[\psi,\psi]
& = W[ \psi_t,\psi] + W[\psi,\psi_t] \\
& = W[-\alpha |\psi|^2  \psi_x
+ i \beta  \psi_{xx}
- \gamma  \psi_{xxx}
+i \theta |\psi|^2 \psi ,\psi] \\
& \quad + W[\psi,-\alpha |\psi|^2 \psi_x
+ i \beta  \psi_{xx}
- \gamma  \psi_{xxx}
+i \theta |\psi|^2 \psi ] \\
& = -\alpha \Big( W[|\psi|^2  \psi_x,\psi] + W[\psi,|\psi|^2 \psi_x] \Big) \\
& \quad + i \beta \Big( W[ \psi_{xx},\psi] - W[\psi, \psi_{xx}] \Big) \\
& \quad - \gamma \Big( W[ \psi_{xxx},\psi] + W[\psi, \psi_{xxx}] \Big) \\
& \quad + i \theta \Big( W[|\psi|^2 \psi,\psi] - W[\psi,|\psi|^2 \psi] \Big). 
\end{align*}
We can go further by appealing to the Wigner-Moyal transform properties, that is
\begin{equation}\label{eq:proposition 6}
W[u,v]=(W[v,u])^{*}, \quad
W[\partial_xu,v]
= \Big( i k + \frac{1}{2} \partial_x \Big)W[u,v],
\end{equation}
which yields the following equalities
\begin{align}
     W[\psi_{xx}, \psi ] - W[\psi, \psi_{xx}] = 2ik W_x[\psi,\psi], \label{first}
\end{align}
and
\begin{align}
    W[\psi_{xxx}, \psi] + W[\psi, \psi_{xxx}]  =-3k^2 W_{x} + \frac{1}{4} W_{xxx}[\psi,\psi].\label{second}
\end{align}
Using the property of the convolution of Fourier transforms we further obtain
\begin{align}
&W[|\psi|^2 \psi, \psi] - W[\psi, |\psi|^2 \psi] \nonumber \\ 
& \quad = \frac{1}{2\pi} \int_{\R}\int_{\R}e^{-i\lambda y} \Big\{ Z(x+y/2, t)  - Z(x-y/2, t)\Big\} dy \label{third}\\
& \qquad \qquad \qquad \times W[\psi,\psi](x, k-\lambda, t) d\lambda, \nonumber 
\end{align}
and 
\begin{align}
    & W[|\psi|^2 \psi_x, \psi] + W[\psi, |\psi|^2 \psi_x] \nonumber \\ & = \int_{\R}\int_{\R} \frac{i (k-\lambda)e^{- i \lambda y}}{2\pi}   \, 
\Big\{ Z(x+\frac{y}{2},t) - Z(x-\frac{y}{2},t) \Big\}dy \nonumber\\  
& \qquad \qquad \qquad  \times W[\psi, \psi](x,k-\lambda,t) d\lambda \label{fourth}\\
& + \int_{\R}\int_{\R} \frac{e^{- i \lambda y}}{4\pi}  \, 
\Big\{ Z\Big(x+\frac{y}{2},t\Big) + Z\Big(x-\frac{y}{2},t\Big) \Big\}dy \nonumber \\
&  \qquad \qquad \qquad  \times W_x[\psi, \psi](x,k-\lambda,t) d\lambda , \nonumber
\end{align}
where in the last expression we again used \eqref{eq:proposition 6}. Now, after gathering \eqref{first}, \eqref{second}, \eqref{third}, \eqref{fourth}, we arrive at Wigner-Hirota equation
\eqref{eq:HirotaWigner}.
\section{The dispersion relation }\label{App2}

Here we show how to obtain the dispersion relation (\ref{eq:dispersion relation}), as well as the derivation of the expressions (\ref{eq:growthrate}) and (\ref{Omega Lorentz}). By introducing the ansatz \eqref{eq:ansatz} into Eq. \eqref{eq:HirotaWigner}, we obtain the following 
\begin{align*}
& -i  \varepsilon \, F_{K,\Omega}(k) \, \Omega  \, 
  e^{i(Kx - \Omega t)}
 =  \Big(-2\beta k + 3\gamma k^2 \Big) i \varepsilon \, F_{K,\Omega}(k) \, K \,  e^{i(Kx - \Omega t)}  \\ 
& + \frac{1}{4}\gamma  \, i \varepsilon \, F_{K,\Omega}(k) \, K^3 \,  e^{i(Kx - \Omega t)} \\
&  + \frac{1}{2\pi} \int_{\R}\int_{\R} \Big(i \theta - i \alpha (k-\lambda) \Big)e^{-i \lambda y} 
\Big\{ Z\Big(x+\frac{y}{2},t\Big) - Z\Big(x-\frac{y}{2},t\Big) \Big\} \, dy \\ 
& \qquad \times  \Big( W_0(k-\lambda) + \varepsilon \,  F_{K,\Omega}(k-\lambda) \,e^{i(Kx - \Omega t)} \Big) d\lambda \\
&  -  \frac{\alpha}{4\pi} i \varepsilon \, K \,  e^{i(Kx - \Omega t)} \\
&\qquad \times  \int_{\R}\int_{\R}  e^{-i \lambda y}
\Big\{ Z\Big(x+\frac{y}{2},t\Big) + Z\Big(x-\frac{y}{2},t\Big) \Big\}\, F_{K,\Omega}(k-\lambda)\, dy \, d\lambda,
\end{align*}
where
\begin{align*}
Z(x,t)
= \frac{1}{2\pi}\int_{\mathbb{R}}  W_0(k) \, dk + \frac{1}{2\pi} \varepsilon  \, e^{i(Kx - \Omega t)}\int_{\mathbb{R}}\, F_{K,\Omega}(k) \, dk.
\end{align*}

The above formula has higher order terms in $\varepsilon$, which we neglect in agreement with the MI analysis. The linearized version is as follows
\begin{align*}
& - \,F_{K,\Omega}(k) \, \Omega 
 =  \Big(- 2\beta  k    + 3\gamma  k^2 \Big)  \,F_{K,\Omega}(k) \, K 
 + \frac{1}{4}\gamma  \, F_{K,\Omega}(k)   \, K^3 \,  \nonumber \\
 & - \frac{1}{2\pi} \alpha \,F_{K,\Omega}(k) \, K \, \psi_0^2+ \frac{ i}{4\pi^2}\int_{\R}F_{K,\Omega}(k) \, dk \, \\
& \times \int_{\R}\Big(\theta - \alpha  (k-\lambda) \Big) 
 W_0(k-\lambda) \int_{\R} e^{- i \lambda y} \, 
\sin\bigg(\frac{Ky}{2}\bigg) \, dy  \,   d\lambda .\nonumber
\end{align*}
After plugging the Fourier transform of the sine function in the right hand side and reorganizing the expression we obtain
\begin{align*}
&\frac{F_{K,\Omega}(k)}{\int_\R F_{K,\Omega}(k) dk} \\  
& \qquad =
-\frac{1}{4\pi} \Big((\alpha k - \theta)\big(W_0(k+K/2)-W_0(k-K/2)\big) \\
&+\alpha \frac{K}{2}\big(W_0(k+K/2)+W_0(k-K/2)\big)\Big)/\Big(\Omega+\frac{1}{4}\gamma  \,  K^3  - \frac{1}{2\pi}\alpha K\psi_0^2 \\
&+\big(- 2\beta k + 3\gamma  k^2 \big) \, K \Big).\end{align*}
Finally, integrating in $k$, the above formula can be rewritten as
\begin{eqnarray}
1&+&\frac{1}{4\pi}
\int_{\R}\Big(\Omega
+\frac{\gamma}{4} K^3
- \frac{\alpha \psi_0^2}{2\pi}K
+\big(- 2\beta k + 3\gamma  k^2 \big) \, K\Big)^{-1} \nonumber \\ 
&\times& \Big( (\alpha k - \theta)\big(W_0(k+K/2)-W_0(k-K/2)\big) \nonumber  \\
&+&\alpha \frac{K}{2}\big(W_0(k+K/2)+W_0(k-K/2)\big)\Big) dk=0. \nonumber
\end{eqnarray}
which, yields the dispersion relation (\ref{eq:dispersion relation}), as wanted. 

Accordingly, the integral involved in the dispersion relation can be performed by means of the standard contour-integration techniques. Clearly, one of the poles is determined from the roots of the polynomial in the denominator, that is
\begin{equation}
\Omega +\frac{\gamma}{4} K^3 - \frac{\alpha \psi_0^2}{2\pi}K +\big(- 2\beta k + 3\gamma  k^2 \big) \, K=0,
\label{Aroots0}
\end{equation}
which are given by
\begin{equation}
 k_{\pm}=\frac{1}{6 \gamma }\Bigg( 2 \beta \pm\sqrt{4 \beta ^2-\frac{3 \gamma  \left(\pi  \gamma  K^3-2 \alpha K \psi_{0}^2+4 \pi  \Omega \right)}{\pi }}\Bigg) . 
 \label{Aroots1}
\end{equation}
As mentioned in the Sec.\ref{WIA}, these poles may switch between the upper or lower half plane for the broad set of complex values of the modulation frequency $\Omega$ and the problem parameters, which prevents to discern a general form of the residues for a selected integration contour. Nonetheless, in the non-dispersive situation $\gamma=0$ this boils down to a pole whose location is fully determined by $\Omega$, i.e. 
\begin{equation}
k_{\pm}\rightarrow  k_{0}= \frac{1}{4 \pi  \beta }\Big(2 \pi  \Omega -\alpha  K \psi_{0}^2 \Big).
\label{Aroots2}
\end{equation}

\subsection{Benjamin-Feir instability}

In this appendix we show how to obtain the dispersion relation (\ref{eq:growthrate}) for the stationary solution (\ref{eq: deltaWigner}).
Inserting the latter into the equation \eqref{eq:dispersion relation} yields
\begin{align}
0
& = 1+\frac{\psi_0^2}{4\pi}
\int_{\R}
\frac{1}{
\Omega
+\frac{\gamma}{4} K^3
- \frac{\alpha \psi_0^2}{2\pi}K
+\big(- 2\beta k + 3\gamma  k^2 \big) \, K } \nonumber \\
&  \quad \times \Big\{ (\alpha k - \theta)\big(\delta(k+K/2)- \delta(k-K/2)\big) \nonumber  \\
& \qquad +\alpha \frac{K}{2}\big(\delta(k+K/2) + \delta(k-K/2)\big) \Big\} \, dk \nonumber.
\end{align}
By separating the integral into two parts weighted by Dirac delta functions, we immediately obtain the following algebraic expression
\begin{align*}
1+ \frac{\psi_0^2}{4\pi} \Big[ 
& -\frac{\theta}{\Omega + \gamma K^3 + \beta K^2 - \alpha \psi_0^2 K /2\pi} \\
& + \frac{\theta}{\Omega + \gamma K^3 - \beta K^2 - \alpha \psi_0^2 K /2\pi}\Big] = 0 .
\end{align*}
The desired dispersion relation (\ref{eq:growthrate}) is directly obtained by solving this equation in terms of $\Omega$.

\subsection{Incoherent modulation instability: instability growth rate}

We now present the derivation of the relation (\ref{Omega Lorentz}) associated to the Lorentzian-type stationary solution when the TOD can be neglected (i.e. $\gamma=0$). By replacing \eqref{LW0} into the general dispersion relation \eqref{eq:dispersion relation}, we arrive to the expression 
\begin{align*}
    & 1+ \frac{\psi_0^2 K p_0}{4\pi^2} \int_{\R} \big(-\alpha  k + 2\theta k + \alpha (p_0^2 + K^2/4)\big) \\
    & \quad \times \Big[ (3\gamma K k^2 - 2\beta K k + \Omega + \gamma K^3/4 - \alpha K \psi_0^2/2\pi) \Big]^{-1} \\
    & \quad \times \Big[ \big( k-(ip_0 + \frac{K}{2}) \big)\big(k - (ip_0 - \frac{K}{2})\big)\Big]^{-1}\\
    & \quad \times \Big[ \big(k-(-ip_0 + \frac{K}{2})\big)\big(k-(-ip_0 - \frac{K}{2})\big) \Big]^{-1} dk = 0.
\end{align*}
As mentioned in Appendix B and Sec. \ref{WIA}, we neglect TOD to simplify computations. Thus, dispersion relations can be rewritten as
\begin{eqnarray}\label{App:CRT}
 1 &+& \frac{\psi_0^2 p_0}{8\beta \pi^2}\int_{\R} \big( \alpha k - 2\theta k - \alpha (p_0^2 + K^2/4) \big) \\
 &\times& \Big[ \big(k-C\big)  \big(k-(ip_0 + \frac{K}{2})\big)\big(k-(ip_0 - \frac{K}{2})\big) \Big]^{-1} \nonumber \\
 &\times& \Big[ \big( k - (-ip_0 + \frac{K}{2})\big)\big(k-(-ip_0 -\frac{K}{2})\big) \Big]^{-1} dk =0, \nonumber
\end{eqnarray}
where $C = \frac{\Omega}{2\beta K} -\frac{\alpha\psi_0^2}{4\beta\pi}$. Notice that $C$ is proportional to the root given by the Eq.(\ref{Aroots2}) mentioned in the previous appendix, i.e. $C=k_{0}/K$.

As stated before the integration (\ref{App:CRT}) can be carried out by using contour integration techniques. So, in this integration we find 5 poles: 
2 with negative imaginary part, 
2 with positive imaginary part 
and one pole $C$. We set $Im(C)<0$ and perform the contour integration upon the upper-half complex plane for simplicity, which means to assume $K<0$. As a result we obtain a quadratic equation in $\Omega$
\begin{align}\label{App:CRT1}
    & \Big( \frac{\Omega}{2\beta K} - \frac{\alpha \psi_0^2}{4\beta \pi} \Big)^2 -2ip_0 \Big(\frac{\Omega}{2\beta K} - \frac{\alpha \psi_0^2}{4\beta \pi}\Big) \\
    & \quad \quad - \frac{K^2}{4} -p_0^2 - \frac{\alpha \psi_0^2}{8\pi \beta} ip_0 + \frac{\theta \psi_0^2}{8\pi \beta}=0,\nonumber
\end{align}
which we solve to find \eqref{Omega Lorentz}.  Recall that if $\gamma \neq 0$ we would have six roots to deal with, since the fifth pole (denoted by $C$) would be replaced by the two poles determined by the roots (\ref{Aroots1}). Although we could proceed in the same way, it would not be straightforward to apply the residue theorem as mentioned in the previous appendix (see discussion below the Eq.(\ref{Aroots1})).\\

\subsection{Incoherent modulation instability: unstable wavenumber set}

Accordingly, in order to deduce \eqref{Eq:SPC}, we look for those $K$ in \eqref{Omega Lorentz} such that $Im(\Omega)>0$.  Using the polar representation, we can write the square root expression in \eqref{Omega Lorentz} as
\begin{align*}
    \sqrt{r} \cos{(\mu/2)} + i\sqrt{r}\sin{(\mu/2)}, 
\end{align*}
where $$r = \sqrt{a^2 + b^2}, \qquad \mu = \tan^{-1} (b/a) ,$$
with $$ a = \frac{\psi_0^2 \theta}{2\pi\beta K^2} -1, \qquad b = -\frac{\psi_0^2}{2\pi\beta K^2} p_0 \alpha.$$

Thus, for the instability analysis it is enough to check that
\begin{equation}\label{Imaginary Lorentz}
    Im(\Omega) = 2p_0 \beta K \pm \beta K^2 \sqrt{r} \cos{(\mu/2)} >0.
\end{equation}

We start analyzing the expression $\cos{(\mu/2)}$.
For $\mu \in [0, 2\pi)$ we have that
$$\frac{b}{2a} = \frac{-\frac{\psi_0^2}{2\pi\beta K^2} p_0\alpha}{\frac{\psi_0^2}{\pi\beta K^2} \theta -2}.$$
Since we fixed all parameters in the Hirota equation to be positive, we observe that $b<0$. Furthermore, it is convenient to focus first on the situation of $K<0$, and consider the cases $a<0$ and $a>0$ separately.\\

Assume first that $a<0$, which means 
\begin{equation}\label{firstcase}
    K \in \Big(-\infty, -\sqrt{\frac{\psi_0^2 \theta}{2\pi\beta}}\Big).
\end{equation}
Also, $a<0$ and $b<0$ implies $\mu \in (\pi, 3\pi/2)$, and $\mu/2 \in (\pi/2, 3\pi/4)$. Hence, recalling simple trigonometric properties, we have that
\begin{align*}
&  \sqrt{r} \cos{(\mu/2)} 
 = -\sqrt{r} \sqrt{\frac{1+\cos{\mu}}{2}} \\
& \qquad  = -\sqrt{\frac{r}{2}}\sqrt{1-\frac{1}{\sqrt{1+ (b/a)^2}}} = -\sqrt{\frac{r}{2}}\sqrt{1-\frac{|a|}{r}} \\
 & \qquad = -\sqrt{\frac{r}{2}} \sqrt{\frac{r-|a|}{r}}  = -\frac{1}{\sqrt{2}} \sqrt{r-|a|} = -\frac{1}{\sqrt{2}} \sqrt{r+a}.
\end{align*}

Therefore, the condition \eqref{Imaginary Lorentz} can be rewritten as follows
\begin{align*}
    Im(\Omega) 
    & = \pm \frac{\beta K^2}{\sqrt{2}} \sqrt{r+a} + 2p_0 \beta K\\
    & = \beta K(\pm \frac{K}{\sqrt{2}} \sqrt{r+a} + 2p_0) >0.
\end{align*}
Since $K<0$, we need that
$$\frac{K}{\sqrt{2}} \sqrt{\sqrt{a^2+b^2}+a}+2p_0 <0,$$
or equivalently,
$$ \frac{K^2}{2} (\sqrt{a^2+b^2} + a) > 4p_0^2.$$
Substituting back expressions for $a,b$ we deduce
$$K^2 < \frac{\psi_0^2 \theta}{2\pi\beta}+ \bigg(\frac{\psi_0^2 \alpha}{8\pi\beta}\bigg)^2-4p_0^2,$$
that is,
$$K > -\sqrt{ \frac{\psi_0^2 \theta}{2\pi\beta}+ \bigg(\frac{\psi_0^2 \alpha}{8\pi\beta}\bigg)^2-4p_0^2}.$$

Combining with \eqref{firstcase} we obtain the unstable wavenumber set
\begin{equation*}
  K \in \Bigg[ -\sqrt{\frac{\psi_0^2 \theta}{2\pi\beta}+ \bigg(\frac{\psi_0^2 \alpha}{8\pi\beta}\bigg)^2-4p_0^2} \,\, , -\sqrt{\frac{\psi_0^2\theta}{2\pi\beta}}\Bigg].  
\end{equation*}

Second we treat the situation $a>0$.
Since, $K<0$ we have the following interval for K
\begin{equation}\label{secondcase}
    K\in \Big[- \sqrt{\frac{\psi_0^2 \theta}{2\pi\beta}}, 0 \Big].
\end{equation}

Now, $a>0$ and $b<0$, so we have that $\mu \in (3\pi/2, 2\pi)$ and $\mu \in (3\pi/4, \pi)$. Using again trigonometric properties, we find that
$$\sqrt{r} \cos{(\mu/2)} = -\frac{1}{\sqrt{2}} \sqrt{r+a}.$$
This is exactly the same as in the case of $a<0$, and the computations for the instability interval leads to similar conclusions
$$K > -\sqrt{\frac{\psi_0^2 \theta}{2\pi\beta}+ \bigg(\frac{\psi_0^2 \alpha}{8\pi\beta}\bigg)^2-4p_0^2}.$$
However, with \eqref{secondcase} we have that
\begin{equation*}
K\in \Bigg[ -\sqrt{\frac{\psi_0^2 \theta}{2\pi\beta}} \,\, , 0 \Bigg].
\end{equation*}

Combining both cases, we obtain the unstable wavenumber set for $K<0$:
\begin{equation}\label{intervalfinal}
K\in \Bigg[ -\sqrt{ \frac{\psi_0^2 \theta}{2\pi\beta}+ \bigg(\frac{\psi_0^2 \alpha}{8\pi\beta}\bigg)^2-4p_0^2} \,\,, 0\Bigg].
\end{equation}

The instability interval for $K>0$ can be also obtained from the Eq.(\ref{intervalfinal}) by appealing to the fact that the contour integration in (\ref{App:CRT}) when it is performed upon the upper-half complex plane returns an identical result (\ref{App:CRT1}) by replacing the poles by its complex conjugate instead, i.e. $Im(C^{*}) = Im(\Omega)/2\beta (-K)$. As a consequence, the result (\ref{intervalfinal}) must hold when taking $-K$ as well, which immediately implies \eqref{Eq:SPC}, as we wanted to show.

\newpage
%
\end{document}